\documentclass[twocolumn]{aastex63}

\usepackage{xcolor,relsize}

\shorttitle{The Quasar Dipole}
\shortauthors{Secrest et al.}

\begin{document}

\title{A Test of the Cosmological Principle with Quasars}

\begin{abstract}
We study the large-scale anisotropy of the Universe by measuring 
the dipole in the angular distribution of a flux-limited, all-sky sample of 1.36~million quasars observed by the Wide-field Infrared Survey Explorer (WISE). This sample is derived from the new CatWISE2020 catalog, which contains deep photometric measurements at 3.4 and 4.6~\micron\ from the cryogenic, post-cryogenic, and reactivation phases of the WISE mission. While the direction of the dipole in the quasar sky is similar to that of the cosmic microwave background (CMB), its amplitude is over twice as large as expected, rejecting the canonical, exclusively kinematic interpretation of the CMB dipole with a p-value of $5\times10^{-7}$ ($4.9\sigma$ for a normal distribution, one-sided), the highest  significance achieved to date in such studies. Our results are in conflict with the cosmological principle, a foundational assumption of the concordance $\Lambda$CDM model.
\end{abstract}

\correspondingauthor{Nathan\ J.\ Secrest}
\email{nathan.j.secrest.civ@mail.mil}

\author[0000-0002-4902-8077]{Nathan J.\ Secrest}
\affiliation{U.S.\ Naval Observatory, 3450 Massachusetts Ave NW, Washington, DC 20392-5420, USA}

\author[0000-0002-6274-1424]{Sebastian von Hausegger}
\affiliation{INRIA, 615 Rue du Jardin-Botanique, 54600 Nancy Grand-Est, France}
\affiliation{Sorbonne Universit\'e, CNRS, Institut d'Astrophysique de Paris, 98bis Bld Arago, Paris 75014, France}
\affiliation{Rudolf Peierls Centre for Theoretical Physics, University of Oxford, Parks Road, Oxford, OX1 3PU, United Kingdom}

\author[0000-0001-5023-5631]{Mohamed Rameez}
\affiliation{Dept. of High Energy Physics, Tata Institute of Fundamental Research, Homi Bhabha Road, Mumbai 400005, India}

\author[0000-0002-5944-3995]{Roya Mohayaee}
\affiliation{Sorbonne Universit\'e, CNRS, Institut d'Astrophysique de Paris, 98bis Bld Arago, Paris 75014, France}

\author[0000-0002-3542-858X]{Subir Sarkar}
\affiliation{Rudolf Peierls Centre for Theoretical Physics, University of Oxford, Parks Road, Oxford, OX1 3PU, United Kingdom}

\author[0000-0003-3300-2507]{Jacques Colin}
\affiliation{Sorbonne Universit\'e, CNRS, Institut d'Astrophysique de Paris, 98bis Bld Arago, Paris 75014, France}


\section{Introduction}
The standard Friedmann-Lema\^{i}tre-Robertson-Walker (FLRW) cosmology is based on the ``cosmological principle'', which posits that the universe is homogeneous and isotropic on large scales. This assumption is supported by the smoothness of the CMB,  which has temperature fluctuations of only $\sim 1$ part in 100,000 on small angular scales. These higher multipoles of the CMB angular power spectrum are attributed to Gaussian density fluctuations created in the early universe with a nearly scale-invariant spectrum, which have grown through gravitational instability to create the large-scale structure in the present universe. The dipole anisotropy of the CMB is however much larger, being about 1 part in 1000 as observed in the heliocentric rest frame. This is interpreted as due to our motion with respect to the rest frame in which the CMB is isotropic, and is thus called the kinematic dipole. According to the most recent measurements, the inferred  velocity is $369.82 \pm 0.11$~km\,s$^{-1}$ towards $l,b=264\fdg021,48\fdg253$  \citep{2018arXiv180706205P}. This motion is usually attributed to the gravitational effect of the inhomogeneous distribution of matter on local scales, originally dubbed the ``Great Attractor'' \citep[see, e.g.,][]{1991Natur.350..391D}.

A consistency check of the above kinematic interpretation of the CMB dipole would be to measure the concomitant effects on higher multipoles in the CMB angular power spectrum \citep{2002PhRvD..65j3001C}. However, even the precise measurements of these by Planck allow up to 40\% of the observed dipole to be due to effects other than the Solar System's motion \citep[see discussion in][]{2016CQGra..33r4001S}. According to galaxy counts in large-scale surveys, the universe is sensibly homogeneous when averaged over scales larger than $\gtrsim100$~Mpc, as is indeed expected from  considerations of structure formation in the concordance $\Lambda$CDM model. Hence the reference frame of matter at still greater distances should converge to that of the CMB; i.e.\ the dipole in the distribution of cosmologically distant sources, induced by our motion via special relativistic aberration and Doppler shifting effects, should align both in direction and in amplitude with the CMB dipole. Independent measurements of the distant matter dipole are therefore a crucial test of the cosmological principle, and equivalently of the standard model of cosmology.

\cite{1984MNRAS.206..377E} proposed that such a test be done using counts of radio sources. These are typically active galactic nuclei (AGNs) at moderate redshift ($z\sim1$), so locally clustered sources ($z<0.1$), which can introduce an additional dipole in the distribution of matter \citep[e.g.,][]{2016JCAP...03..062T}, are not a significant contaminant. Consider a population of sources with power-law spectra $S_\nu \propto \nu^{-\alpha}$, and integral source counts per unit solid angle $\mathrm{d}N/\mathrm{d}\Omega\,(>S_\nu) \propto S_\nu^{-x}$, above some limiting flux density $S_\nu$. If we are moving with velocity $v\ll c$ with respect to the frame in which these sources are isotropically distributed, then being ``tilted observers'' we should see a dipole anisotropy of amplitude \citep{1984MNRAS.206..377E}:

\begin{equation}
   \mathcal{D} = [2 + x(1+\alpha)]v/c .
   \label{eq:D}
\end{equation}

The advent of the 1.4~GHz NRAO VLA Sky Survey \citep[NVSS;][]{1998AJ....115.1693C}, which contains $\sim1.8$~million sources, enabled the  first estimates of the matter dipole anisotropy \citep{2002Natur.416..150B,2011ApJ...742L..23S,2012MNRAS.427.1994G,2015APh....61....1T,2015MNRAS.447.2658T,2016JCAP...03..062T}. To improve sky coverage, data was  added from other radio surveys, e.g.\ the 325~MHz Westerbork Northern Sky Survey \citep[WENSS;][]{1997A&AS..124..259R,2013A&A...555A.117R}, the 843~MHz Sydney University Molonglo Sky Survey \citep[SUMMS;][]{2003MNRAS.342.1117M,2017MNRAS.471.1045C,2019ApJ...878...32T} and the 150~MHz TIFR GMRT Sky Survey \citep[TGSS;][]{2018JCAP...04..031B,2019PhRvD.100f3501S}. However, as was first noted by \cite{2011ApJ...742L..23S}, while the direction of the matter dipole is consistent with that of the CMB, its amplitude is several times larger.

In this Letter, we report the first independent measurement of the dipole in the angular distribution of distant quasars using mid-infrared data from the Wide-field Infrared Survey Explorer \citep[WISE;][]{2010AJ....140.1868W}, which surveyed the sky at 3.4\,\micron, 4.6\,\micron, \,12\micron, and 22\,\micron\ (W1, W2, W3, and W4). This provides a measurement of the dipole that is independent of the radio survey-based results, as WISE is a space mission with its own unique scanning pattern, not constrained by the same observational systematics that affect ground-based surveys, such as declination limits or atmospheric effects. While WISE, along with 2MASS, has been used before to set useful constraints on the matter dipole \citep{2012MNRAS.427.1994G,2014MNRAS.445L..60Y,2015MNRAS.449..670A,2017MNRAS.464..768B,2018MNRAS.477.1772R}, these studies were of relatively nearby galaxies ($z\sim0.05-0.1$) where contamination from local sources can be significant and has to be carefully accounted for. In Section~\ref{sec: sample}, we detail the quasar sample that we use, and we introduce our methodology in Section~\ref{section: Method}. Our results are presented in Section~\ref{sec: results}, and we discuss their significance for cosmology in Section~\ref{sec: conclusions}.

\section{Quasar Sample} \label{sec: sample}
Because of the unique power of mid-infrared photometry to pick out AGNs, WISE may be used to create reliable AGN/quasar catalogs based on mid-infrared color alone \citep[e.g.,][]{2015ApJS..221...12S}. We require an AGN sample optimized for cosmological studies, so the objects should preferably be quasars: AGN-dominated and at moderate or high-redshift \citep[$z\gtrsim0.1$; cf.,][]{2016JCAP...03..062T}. 
The sample should cover as much of the celestial sphere as is possible to minimize the impact of missing (or masked) regions, and be as deep as possible to contain the largest number of objects and thus have the greatest statistical power.

We created a custom quasar sample from the new CatWISE2020 data release \citep{2020ApJS..247...69E}, which contains sources from the combined 4-band cryo, 3-band cryo, post-cryo NEOWISE, and reactivation NEOWISE-R data. The CatWISE2020 catalog is 95\% complete down to $\lesssim17.4$~mag in W1 and $\lesssim17.2$~mag in W2, respectively 0.3~mag and 1.5~mag deeper than the previous AllWISE catalog. We select all sources in the CatWISE2020 catalog with valid measurements in W1 and W2, which are the most sensitive to AGN emission \citep[e.g.,][]{2012ApJ...753...30S}. To avoid any potential directional bias from uncorrected Galactic reddening, we corrected the W1 and W2 magnitudes using the \citet{2014A&A...571A..11P} dust map and the extinction coefficients from \citet{2019ApJ...877..116W}. To select quasars, we impose the color cut $\mathrm{W1-W2}\geq0.8$ \citep{2012ApJ...753...30S}, which indicates AGN-dominated emission following a power-law distribution ($S_\nu \propto \nu^{-\alpha}$). This yields a raw sample of 141,698,603 objects.

To remove poor-quality photometry near clumpy and resolved nebulae both in our Galaxy (e.g., planetary nebulae) and in nearby galaxies such as the Magellanic Clouds and Andromeda, we produced masks for the nebulae and used masks of size 6 times the 20~mag~arcsec$^{-2}$ isophotal radii from the 2MASS Large Galaxy Atlas \citep[LGA;][]{2003AJ....125..525J} for the Magellanic Clouds and Andromeda. In the WISE catalogs, there are often image artifacts near bright stars, caused by density suppression in their vicinity.\footnote{\url{http://wise2.ipac.caltech.edu/docs/release/allsky/expsup/sec6\_2.html\#brt\_stars}} We find that circular masks with 2MASS $K$ band-dependent radii $\log_{10}(r~\mathrm{deg}^{-1}) = -0.134\,K - 0.471$ effectively remove these. We removed any remaining areas of poor photometry or artifacts using masks of radius $\leq2\deg$. In all, we masked 291 sky regions, not including the Galactic plane.

\begin{figure*}
    \centering
    \includegraphics[width=\textwidth]{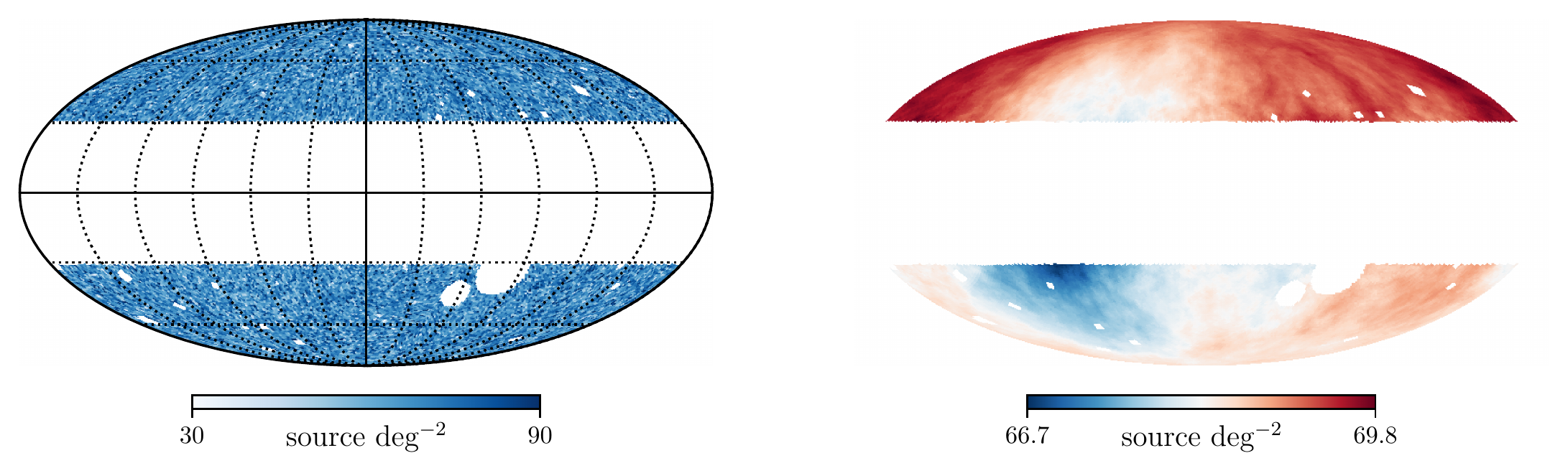}
    \caption{\textit{Left:} Mollweide density map of our CatWISE quasar sample, in Galactic coordinates. \textit{Right:} density map smoothed using a moving average on steradian scales, showing a dipole signal. Both maps have been corrected for the residual ecliptic latitude bias (Section~\ref{sec: sample}).}
    \label{fig:skymaps_final}
\end{figure*}

Above a $W1$ magnitude of 16.4, uneven source density appears due to overlaps in the ecliptic scanning pattern of WISE, most prevalent at the ecliptic poles. We select a magnitude cut of $9>\mathrm{W1}>16.4$ (Vega), where the lower bound guards against potential saturation. While completeness in high source density areas in CatWISE2020 (hereafter CatWISE) is improved over the CatWISE Preliminary Catalog,\footnote{\url{https://catwise.github.io/CatWISE2020_2020_07_18.pdf}} we nonetheless find a drop off in source density below Galactic latitudes of $|b|<30\arcdeg$, and a mild inverse linear trend in source density versus absolute ecliptic latitude. This trend has a slope of $-0.051$ and a zero-latitude intercept of 68.89~deg$^{-2}$. We include this fit as a correction and weighting function in our later calculations. For the Galactic plane, we cut out all sources below $|b|<30\arcdeg$. We also found 57 objects with anomalously low mean coverage depth \texttt{w1cov}~$<80$ that have high values of $W1-W2$. We remove these, leaving a final sample, after masking, of 1,355,352 quasars, as shown in Figure~\ref{fig:skymaps_final}.

\begin{figure}
    \centering
    \includegraphics[width=\columnwidth]{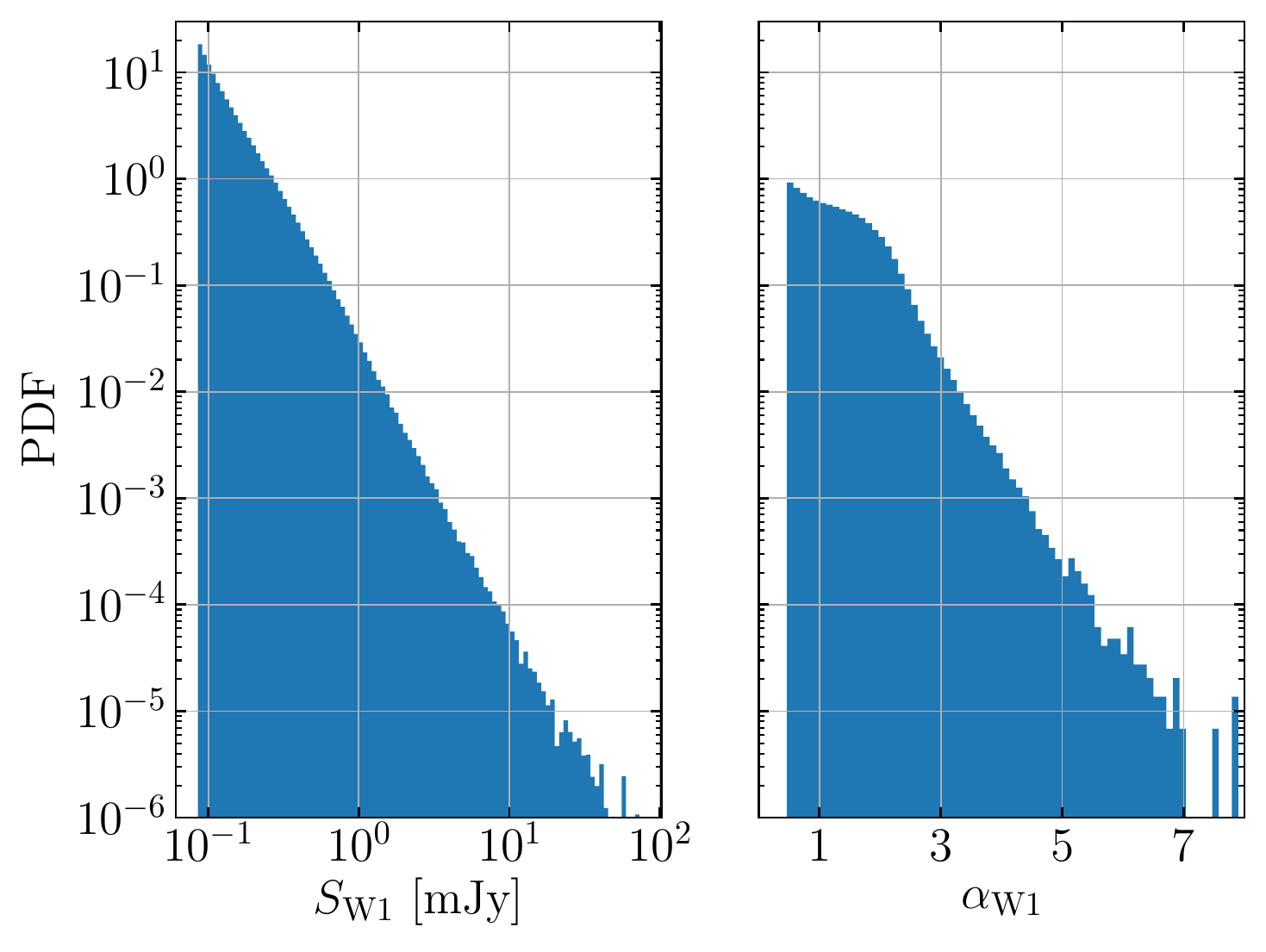}
    \caption{Distribution of flux densities $S_\nu~(\propto \nu^{-\alpha}$) and spectral indices $\alpha$ (W1 band) in our CatWISE quasar sample, normalized as a probability density function (PDF).}
    \label{fig:alphaandflux}
\end{figure}

We calculate spectral indices $\alpha$ of our sample in the W1 band by obtaining power-law fits of the form $S_\nu = k \nu^{-\alpha}$, where $k$ is the normalization. We produced a lookup table to determine $\alpha$ based on $\mathrm{W1-W2}$, by calculating synthetic AB magnitudes following Equation~2 of \citet{2012PASP..124..140B}. The WISE magnitudes are on the Vega magnitude system, so we convert from the AB system using the offsets $m_\mathrm{AB}-m_\mathrm{Vega} = 2.673$, 3.313 for W1 and W2, respectively. These WISE offsets correspond to the constant of $-48.60$ associated with the definition of the synthetic AB magnitude. The normalisation $k$ is calculated by inverting the equation for the synthetic magnitude and using the observed W1 AB magnitude. Finally, we calculate the isophotal frequency, at which the flux density $S_\nu$ equals its mean value within the passband, using Equation~A19 in \citet{2012PASP..124..140B}. As our sample was constructed with the cut $\mathrm{W1-W2}\geq0.8$, the distribution peaks at $\alpha\sim1$ and extends to steeper spectral indices, with a mean value of 1.26. Distributions of spectral indices and fluxes for our final sample of quasars are shown in  Figure~\ref{fig:alphaandflux}. The corresponding mean isophotal frequency is $8.922\times10^{13}$~Hz, with a dispersion of 0.19\%. Our magnitude cut is equivalent to a flux density cut of $77.77>S_\nu>0.09$~mJy.

\begin{figure}
    \centering
    \includegraphics[width=\columnwidth]{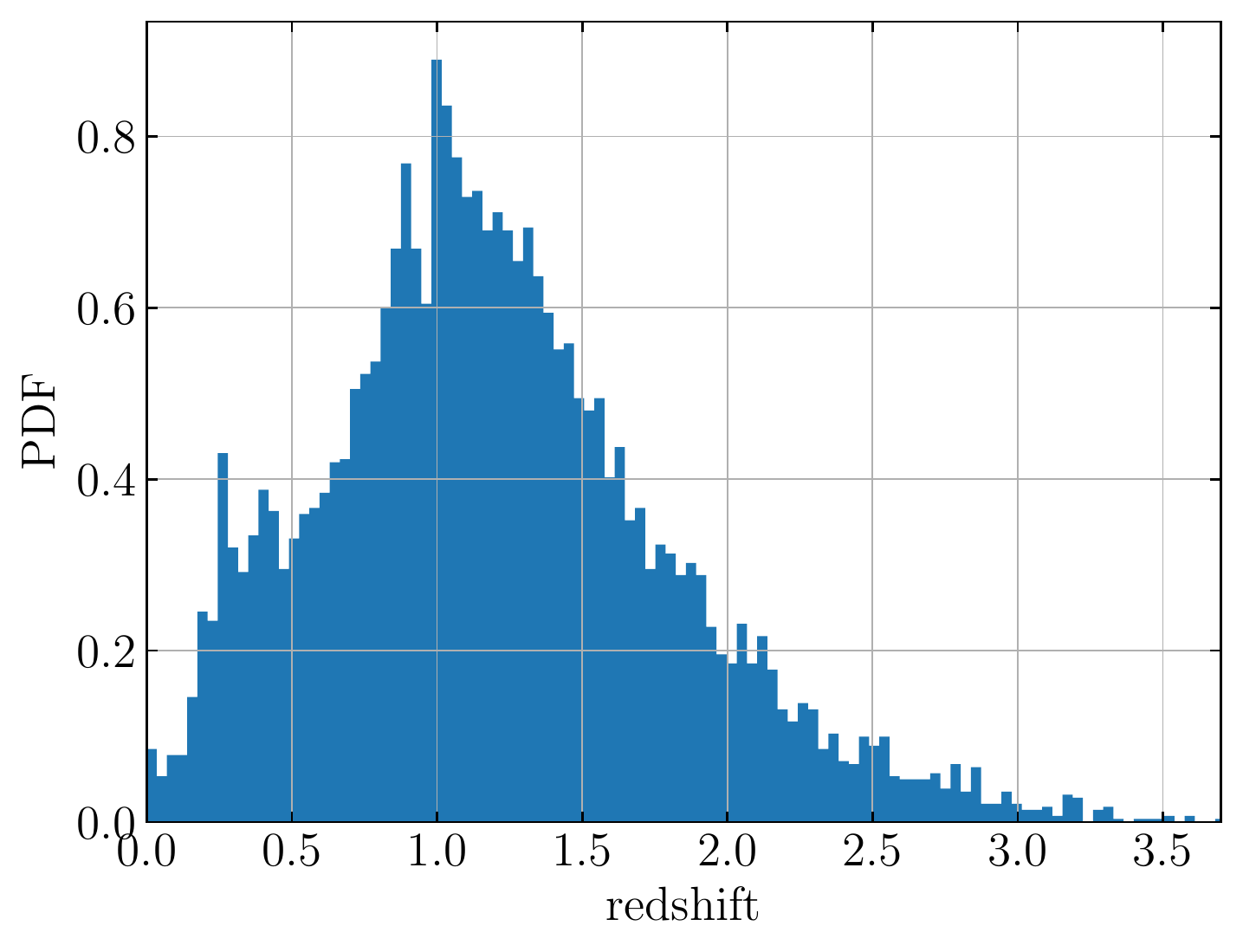}
    \caption{Redshift distribution of our CatWISE quasar sample.}
    \label{fig:zdistro}
\end{figure}

To estimate the distribution of quasar redshifts, we select those within the Sloan Digital Sky Survey (SDSS)~Stripe~82, a 275~deg$^{2}$ region of the sky scanned repeatedly by the SDSS, thus achieving an increase of depth of $\sim2$~mag \citep{2014ApJ...794..120A}. In the \texttt{specObj} table for SDSS~DR16,\footnote{\url{https://www.sdss.org/dr16/spectro/spectro_access}} Stripe~82 contains 4.4 times more objects with spectroscopic $r$-band magnitudes fainter than 20~(AB) than a comparable sky region in the SDSS main footprint. We use a sub-region of Stripe~82 between $-42\arcdeg<\mathrm{R.A.}<45\arcdeg$, which lies outside the $|b|<30\arcdeg$~Galactic plane mask we employ, and which was observed by the Extended~Baryon~Oscillation~Spectroscopic~Survey \citep[eBOSS;][]{2016AJ....151...44D}, yielding even deeper spectral coverage. There are 14,402 CatWISE quasars in this region. For photometric information, we cross-match these with the Dark~Energy~Survey, Data~Release~1 \citep[DES1;][]{2018ApJS..239...18A}, which achieved an $i$-band depth of 23.44~mag (AB). By comparing with Gaia~DR2 \citep{2018A&A...616A...1G}, we found that, in this part of the Stripe~82 footprint, DES1 has a systematic offset in R.A., Decl.\ of $-86$~mas and $+124$~mas. We corrected the DES1 positions for this offset, and use a $10\arcsec$ match for completeness. To avoid spurious matches, we excluded any associations that had a closer counterpart in the CatWISE catalog not in our quasar sample. This produced counterparts for 14,193 quasars (99\%). Matching the DES1 counterpart coordinates onto the \texttt{specObj} table to within $1\arcsec$ for fiber coverage, we find 8594 matches (61\%). The unmatched objects are 0.3 mag fainter in W2 than the matched objects on average, suggesting that they are slightly less luminous or slightly more distant (or both). However, their mean $r-\mathrm{W2}$ value, a measure of AGN obscuration level \citep[e.g.,][]{2013AJ....145...55Y}, is 1.9~mag redder than the mean of the matched sample, implying that the unmatched objects are simply too faint at visual wavelengths for the SDSS. Indeed, while 39\% of the full DES1-matched sample has $r-\mathrm{W2}>6$~mag (Vega), in line with expectations from the literature for the prevalence of type 2 AGNs \citep{2013AJ....145...55Y}, 79\% of the unmatched sample have $r-\mathrm{W2}>6$. This indicates that the objects in our sample without SDSS spectra are predominantly type 2 systems, an effect of the AGN orientation with respect to the line of sight, and so the matched objects may be used to estimate the distribution of redshifts for the full sample. We find a mean redshift of 1.2, with 99\% having $z >  0.1$, so our sample is almost entirely at moderate to high redshift, as shown in Figure~\ref{fig:zdistro}.

\section{Method} \label{section: Method}
\subsection{Dipole Estimator} \label{subsection: estimator}

We determine the dipole $\vec\mathcal{D}$ of our sample using a least-squares estimator:
\begin{equation}
\label{eq:quadest}
\ \mathlarger{\sum}_p \left[n_p - \left(A_0 + \sum_{j=1}^3 A_{1j}d_{j,p}\right)\right]^2,
\label{eq:quadraticestimator}
\end{equation}
where $n_p$ denotes the number density of sources in sky pixel $p$, $A_0$ is the mean density (monopole), $A_{1j}$ are the amplitudes of the three orthogonal dipole templates $d_{j,p}$, and the sum is taken over all unmasked pixels. This expression's analytical minimum with respect to the monopole and dipole amplitudes $A_j$ is found by solving a simple linear equation, as implemented in the \texttt{fit\_dipole} routine of healpy \citep{2019JOSS....4.1298Z}. Using this, the final dipole reads $\vec\mathcal{D} = \left(A_{1,p}/A_0,A_{2,p}/A_0,A_{3,p}/A_0\right)$. We have verified that this estimator does not suffer from bias in either direction or amplitude for density maps simulated in the manner as described below. Before computing the dipole of the source distribution (Figure~\ref{fig:skymaps_final}) the mild inverse linear trend with ecliptic latitude of the source density was taken into account by correcting the latter as described in Section~\ref{sec: sample}.

Similarly to other dipole estimators,~e.g.~\citet{2002Natur.416..150B,2019MNRAS.486.1350B}, our estimator explicitly seeks a dipolar pattern. However, it is neither computationally expensive as the minimization is done analytically, nor prone to leakage into higher multipoles, as it does not force a spherical harmonic decomposition on an incomplete sky.\footnote{Influence from, e.g., a quadrupole on the estimated dipole was found to be negligible.} Estimators that are agnostic with regard to the true underlying signal, such as the linear estimator proposed in, e.g.,~\citet{fisher_lewis_embleton_1987,Crawford:2008nh}, exhibit biases that, while well understood~\citep{2013A&A...555A.117R}, make significance estimations difficult.  Nevertheless, using other estimators, we find general consistency in the dipole direction and amplitude of our sample after having accounted for their biases, as much as is possible.

\subsection{Mock data and statistical significance} \label{sub:mockdata}
We generate mock samples of $N_{\rm init}$ vectors drawn from a statistically isotropic distribution, whose directions are subsequently modified by special relativistic aberration according to an observer boosted with velocity $\vec v$. Each sample is then masked with the same mask that was applied to the data (Figure~\ref{fig:skymaps_final}). In order to respect the exact distribution of flux values and spectral indices in the data, we assign to each simulated source a flux density $S_\nu$ and a spectral index $\alpha$ drawn at random from their empirical distributions (Figure~\ref{fig:alphaandflux}). The sampled fluxes are then modulated depending on source position, $\vec v$, and $\alpha$. Finally, only sources with $S_\nu>S_{\nu, \rm cut}$ are retained, and the number of remaining sources is reduced to the size of the true sample through random selection.

Under the null hypothesis that the measured dipole $\vec \mathcal{D}$ is a consequence of our motion with respect to a frame shared by both quasars and the CMB, we generate a set of mock skies according to the above recipe. For each simulated sky we compute and record $\vec \mathcal{D}^{\rm sim}$. The fraction of mock skies with amplitude $\mathcal{D}$ larger than our empirical sample, and with angular distance from the CMB dipole closer than our sample, gives the p-value with which the null hypothesis is rejected. Note that the effect on our results of the distributions of flux and spectral index (Figure~\ref{fig:alphaandflux}) is included automatically via the bootstrap approach employed for our simulations.

\section{Results} \label{sec: results}
\begin{figure*}
    \centering
    \includegraphics[height=0.21\textheight]{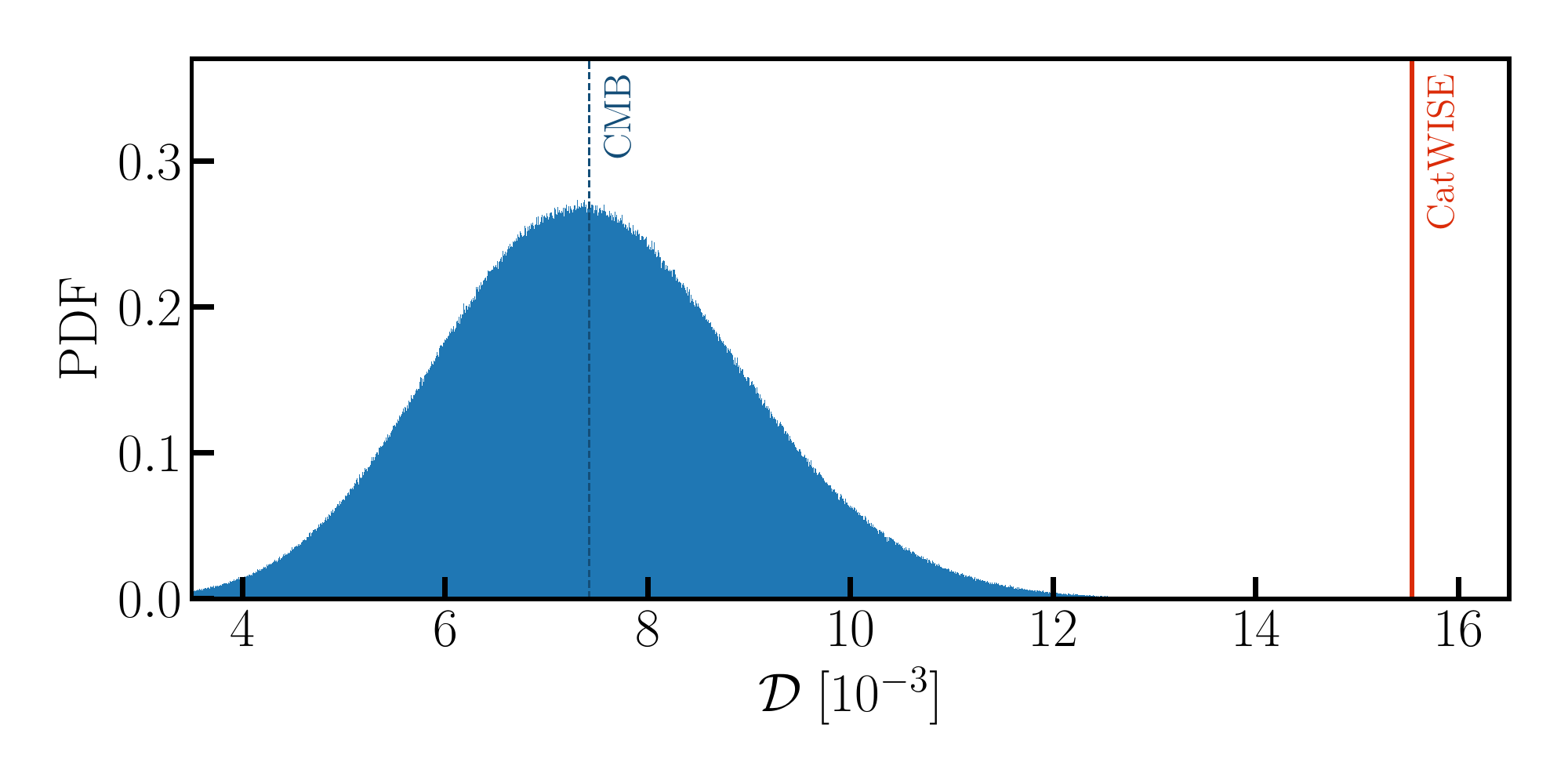}
    \includegraphics[height=0.21\textheight]{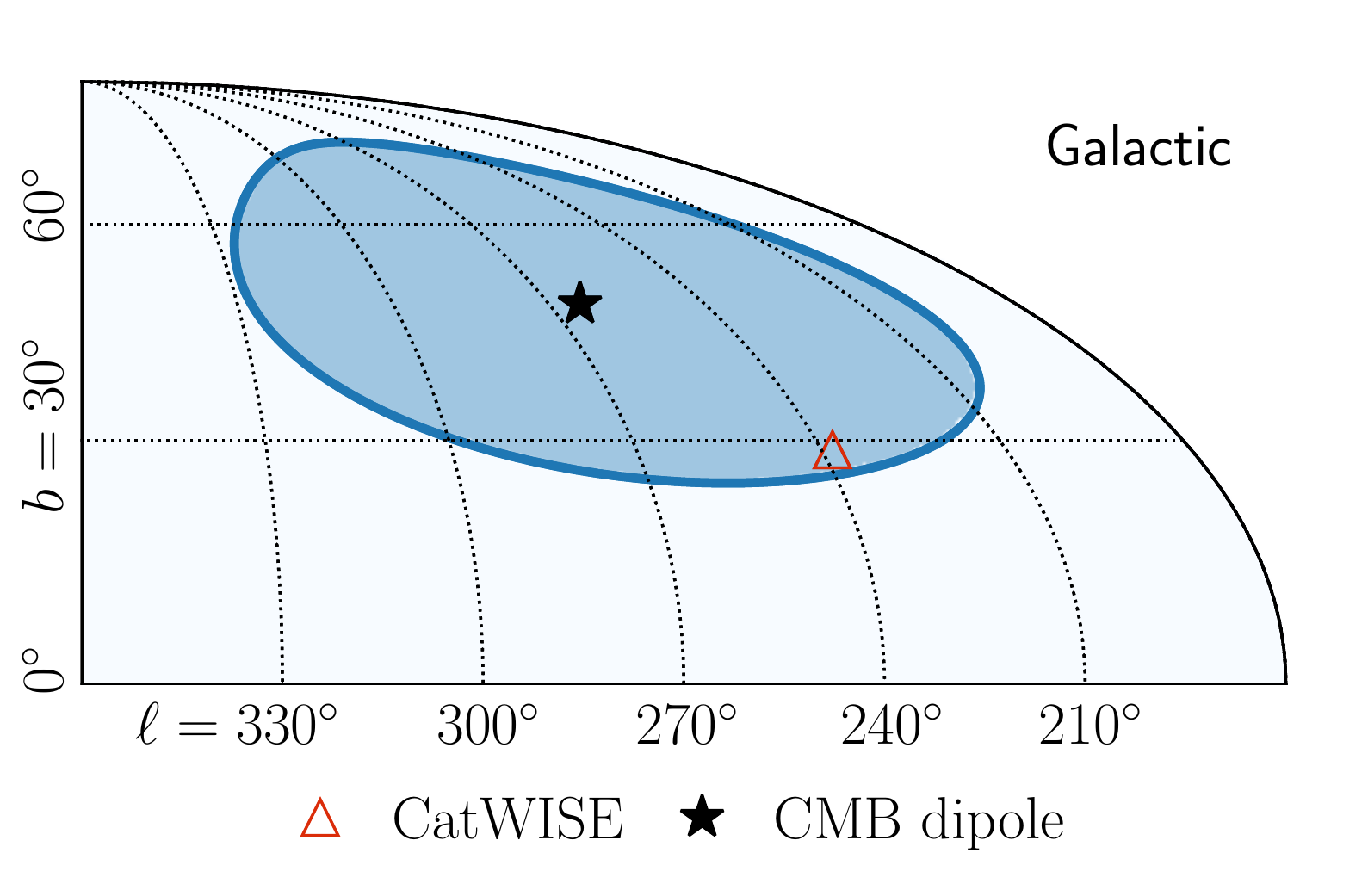}
    \caption{\textit{Left panel:} Amplitude of the dipole $\mathcal{D}$ (solid vertical line) in the CatWISE quasar sample, versus the expectation assuming the kinematic interpretation of the CMB dipole; the distribution of $\mathcal{D}^{\rm sim}$ from simulations (Section~\ref{sub:mockdata}) is shown along with its median value (dashed vertical line). \textit{Right panel:} Dipole direction $\vec\mathcal{D}$ in Galactic coordinates (triangle), with the null hypothesis uncertainty region ($2\sigma$) in blue Section~\ref{sec: results}. The probability under the null hypothesis of observing the dipole that we find is $5\times10^{-7}$, or $4.9\sigma$ for a normal distribution (one-sided).}
    \label{fig:simulationsanddirections}
\end{figure*}

Our sample of 1,355,352 quasars exhibits a dipole with amplitude $\mathcal{D}=0.01554$, pointing towards $(l,b)=(238\fdg2, 28\fdg8)$. This is $27\fdg8$ from the direction of the CMB dipole and over twice as large as the expected amplitude of $\sim0.007$, using Equation~\ref{eq:D}. The amplitude and direction are largely unaffected by varying the mask sizes for both estimators. For instance, doubling the size of the point source masks or  introducing a $10\arcdeg$ mask along the super-galactic plane changes the dipole amplitude by less than $\lesssim5\%$, and the direction of the found dipole varies by $\lesssim5\arcdeg$.

When the expected dipole is simulated assuming the kinematic interpretation of the CMB dipole, only five out of $10^7$ such simulations give $\vec\mathcal{D}^{\rm sim}$ with an amplitude larger than the observed value and within the observed angular distance from the CMB dipole (left panel, Figure~\ref{fig:simulationsanddirections}). We can therefore reject the null hypothesis with a p-value of $5\times10^{-7}$, corresponding to a significance of $4.9\sigma$ for a normal distribution (one-sided).

\section{Discussion} \label{sec: conclusions}
The CatWISE quasar sample exhibits an anomalous dipole, oriented  similarly to the CMB dipole but over twice as large. Whereas a ``clustering dipole'' is also expected from correlations in the spatial distribution of the sources, within the concordance model this can be estimated from the distribution of these sources in redshift (see Appendix~\ref{app:cls}) and the expected matter power spectrum. It is smaller than the dipole we observe in these higher redshift quasars by a factor of $\sim 65$. 

The unique statistical power of our study has allowed us to confirm the anomalously large matter dipole suggested in previous work, which used objects selected at a different wavelength (radio), using surveys completely independent of WISE, namely NVSS, WENNS, SUMMS, and TGSS. The ecliptic scanning pattern of WISE has no relationship with the CMB dipole, so there is no reason to suspect that the dipole we measure in the CatWISE quasar sample is an artifact of the survey. 

After \cite{1984MNRAS.206..377E} proposed this important observational test of the cosmological principle, agreement was initially claimed between the dipole anisotropy of the CMB and that of radio sources \citep{2002Natur.416..150B}. If the rest frame of distant quasars is indeed that of the CMB, it would support the consensus that there exists a cosmological standard of rest, related to quantities measured in our heliocentric frame via a local special relativistic boost. This underpins modern cosmology: for example, the observed redshifts of Type~Ia supernovae are routinely transformed to the ``CMB frame''. From this it is deduced that the Hubble expansion rate is accelerating (isotropically), indicating dominance of a cosmological constant, and this has led to today's concordance $\Lambda$CDM model. If the purely kinematic interpretation of the CMB dipole that underpins the above procedure is in fact suspect, then so are the important conclusions that follow from adopting it. In fact, as observed in our heliocentric frame, the inferred acceleration is essentially a dipole aligned approximately with the local bulk flow of galaxies and towards the CMB dipole \citep{2019A&A...631L..13C}, so \emph{cannot} be due to a cosmological constant. 

If it is established that the distribution of distant matter in the large-scale universe does not share  the same reference frame as the CMB, then it will become imperative to ask whether the differential expansion of space produced by nearby nonlinear structures of voids and walls and filaments can indeed be reduced to just a local boost \citep{2013PhRvD..88h3529W}. Alternatively, the CMB dipole may need to be interpreted in terms of new physics, e.g.\ as a remnant of the pre-inflationary universe \citep{1991PhRvD..44.3737T}. \cite{1988ASPC....4..344G} noted that this issue is closely related to the bulk flow observed in the local universe, which in fact extends out much further than is expected in the concordance $\Lambda$CDM model \citep[e.g.,][]{2011MNRAS.414..264C,2013A&A...560A..90F}. Further work is needed to clarify these important issues. 

As \cite{1984MNRAS.206..377E} emphasized, a serious disagreement between the standards of rest defined by distant quasars and the CMB may require abandoning the standard FLRW cosmology itself. The importance of the test we have carried out can thus not be overstated.

\acknowledgments
We thank the anonymous referee for their insightful review that greatly improved this work. We also thank Jean Souchay for discussions that helped motivate this work, Steph LaMassa for helpful suggestions on Stripe~82, and Wilbur Venus for thoughtful comments. N.J.S., M.R. and S.S.\ gratefully acknowledge the hospitality of the Institut d'Astrophysique de Paris. S.v.H.\ is supported by the EXPLORAGRAM Inria AeX grant and by the Carlsberg Foundation with grant CF19\_0456. The authors made use of \texttt{dustmaps} \citep{2018JOSS....3..695M} to calculate Galactic reddening. The data and software to reproduce the analysis and plots in this Letter can be found at \url{https://doi.org/10.5281/zenodo.4431089}.

\facilities{WISE, Blanco, Sloan}

\software{Astropy \citep{2013A&A...558A..33A, 2018AJ....156..123A}, healpy \citep{2019JOSS....4.1298Z}}


\bibliography{manuscript}

\appendix

\section{Clustering dipole within the concordance model} \label{app:cls}
The clustering dipole $\mathcal{D}_\mathrm{cls}$ in a sample of objects as seen by a typical observer in the concordance $\Lambda$CDM cosmology can be computed given the power spectrum $P(k)$ of (dark) matter density perturbations \citep{2012MNRAS.427.1994G}:

\begin{equation}
    \mathcal{D}_\mathrm{cls} = \sqrt{\frac{9}{4\pi}C_1} ,
\end{equation}

\noindent where

\begin{equation}
    C_l = b^2 \frac{2}{\pi} \int_0^{\infty} f_l(k)^2 P(k)k^2 dk .
\end{equation}

\noindent Here $b$ is the linear bias of the observed objects with respect to the dark matter and the filter function $f_l(k)$ is

\begin{equation}
    f_l(k) = \int_0^{\infty} j_l(kr)f(r)dr ,
\end{equation}

\noindent where $j_l$ is the spherical Bessel function of order $l$ and $f(r)$ is the probability distribution for the comoving distance $r$ to a random object in the survey, given by

\begin{equation}
    f(r) = \frac{H(z)}{H_0 r_0}\frac{dN}{dz} ,
\end{equation}

\noindent normalised such that $\int_0^{\infty} f(r) dr = 1$ and $dN/dz$ is the redshift distribution of the observed objects. Employing $r_0 = c/H_0 = 3000 h^{-1}$~Mpc, Planck~2018 cosmological parameters from Astropy, $P(k)$ at $z=0$ using \textsc{camb} \citep{Lewis:1999bs}, and a cubic-spline fit to the redshift distributions shown in Figure~\ref{fig:zdistro} to determine $dN/dz$, we estimate $\mathcal{D}_\mathrm{cls}$ to be 0.00024 (taking $b=1$) for the CatWISE quasar selection. So, the clustering dipole is quite negligible compared to the observed quasar dipole of $\mathcal{D}=0.01554$ (Section~\ref{sec: results}). 

\end{document}